\documentclass[10pt,twocolumn]{article}

\usepackage[margin={1.8cm,1.8cm}]{geometry}
\usepackage{times}
\usepackage{helvet}
\usepackage{courier}
\usepackage{enumitem}
\usepackage{adjustbox}
\usepackage[numbers,sort]{natbib}
\setlist{nosep}

\usepackage{subfig}
\usepackage{array}

\usepackage{color}
\usepackage{url}
\usepackage{verbatim} 
\usepackage{multirow}
\usepackage{algorithm}
\usepackage[noend]{algorithmic}
\usepackage{xspace}
\usepackage{graphicx}
\usepackage{epsfig}
\usepackage{amsmath}
\usepackage{epstopdf}
\usepackage{amssymb}
\usepackage{pbox}
\usepackage{booktabs}
\newcommand{\eq}[1]{(eq.~\ref{#1})}
\newcommand{\hide}[1]{}
\DeclareMathOperator*{\argmax}{\mathrm{argmax}}

\newcommand{\inner}[2]{\langle #1, #2 \rangle}

\newcommand{\modelname}{\emph{Sceptre}}
\newcommand{\eg}{\emph{e.g.}}
\newcommand{\ie}{\emph{i.e.}}
\newcommand{\modelnamelong}{Substitute and Complementary Edges between Products from Topics in Reviews}

\newcommand{\xhdr}[1]{\vspace{2mm}\noindent{{\bf #1.}}}

\usepackage{authblk}

\graphicspath{{./FIG/}}

\title{
Inferring Networks of Substitutable and Complementary Products
}

\author[1]{Julian McAuley\thanks{jmcauley@ucsd.edu}}
\author[2]{Rahul Pandey\thanks{rahul@pinterest.com}}
\author[2,3]{Jure Leskovec\thanks{jure@cs.stanford.edu}}
\affil[1]{Department of Computer Science, UC San Diego}
\affil[2]{Pinterest}
\affil[3]{Department of Computer Science, Stanford University}

\begin{document}

\maketitle

\begin{abstract}
In a modern recommender system, it is important to understand how products relate to each other. For example, while a user is looking for mobile phones, it might make sense to recommend other phones, but once they \emph{buy} a phone, we might instead want to recommend batteries, cases, or chargers. These two types of recommendations are referred to as \emph{substitutes} and \emph{complements}: substitutes are products that can be purchased \emph{instead of} each other, while complements are products that can be purchased \emph{in addition} to each other.

Here we develop a method to infer networks of substitutable and complementary products. We formulate this as a supervised link prediction task, where we learn the semantics of substitutes and complements from data associated with products.
The primary source of data we use is the text of product reviews, though our method also makes use of features such as ratings, specifications, prices, and brands.
Methodologically, we build topic models that are trained to automatically discover topics from text that are successful at predicting and explaining such relationships. Experimentally, we evaluate our system on the \emph{Amazon} product catalog, a large dataset consisting of 9 million products, 237 million links, and 144 million reviews.
\end{abstract}

\section{Introduction}

Recommender systems are ubiquitous in applications ranging from e-commerce to social media, video, and online news platforms. Such systems help users to navigate a huge selection of items with unprecedented opportunities to meet a variety of special needs and user tastes. Making sense of a large number of products and driving users to new and previously unknown items is key to enhancing user experience and satisfaction~\cite{korenSurvey,netflix,Koren09}.

While most recommender systems focus on analyzing patterns of interest in products to provide personalized recommendations~\cite{korenSurvey,ke,wang11,schein}, another important problem is to understand {\em relationships} between products, in order to surface recommendations that are relevant to a given context~\cite{Linden03,ebay}.
For example, when a user in an online store is examining t-shirts she should receive recommendations for similar t-shirts, or otherwise jeans, sweatshirts, and socks, rather than (say) a movie even though she may very well be interested in it.
From these relationships we can construct a {\em product graph}, where nodes represent products, and edges represent various types of product relationships.  Such product graphs facilitate many important applications: Navigation between related products, discovery of new and previously unknown products, identification of interesting product combinations, and generation of better and more context-relevant recommendations.

Despite the importance of understanding relationships between products there are several interesting questions that make the problem of building product graphs challenging:
What are the common types of relationships we might want to discover? 
What data will allow us to reliably discover relationships between products? 
How do we model the semantics of why certain products are related?---For example, the semantics of why a given t-shirt might be related to a particular pair of jeans are intricate and can only be captured by a highly flexible model.
And finally, how do we scale-up our methods to handle graphs of millions of products and hundreds of millions of relations?

\xhdr{Inferring networks of product relationships}
Here we are interested in inferring networks of relationships between millions of products. Even though our method can be used to learn any type of relationship, we focus on identifying two types of links between products: \emph{substitutes} and \emph{complements} \cite{micro}. \emph{Substitutable} products are those that are inter\-changeable---such as one t-shirt for another, while \emph{complementary} products are those that might be purchased together, such as a t-shirt and jeans.

We design a system titled {\em \modelname{} (\modelnamelong{})}, that is capable of modeling and predicting relationships between products from the text of their reviews and descriptions.
At its core, \modelname{} combines topic modeling and supervised link prediction, by identifying topics in text that are useful as features for predicting links between products. Our model also handles additional features such as brand, price, and rating information, product category information, and allows us to predict multiple types of relations (\eg~substitutes and complements) simultaneously. Moreover, \modelname{} harnesses the fact that
products are arranged in a category hierarchy and allows us to extend this hierarchy to discover `micro-categories'---fine-grained categories of closely related products.

\begin{figure}[t]
\centering
\framebox{\adjustbox{trim={.00\width} {.12\height} {.00\width} {.0\height},clip}{\includegraphics[width=0.75\linewidth]{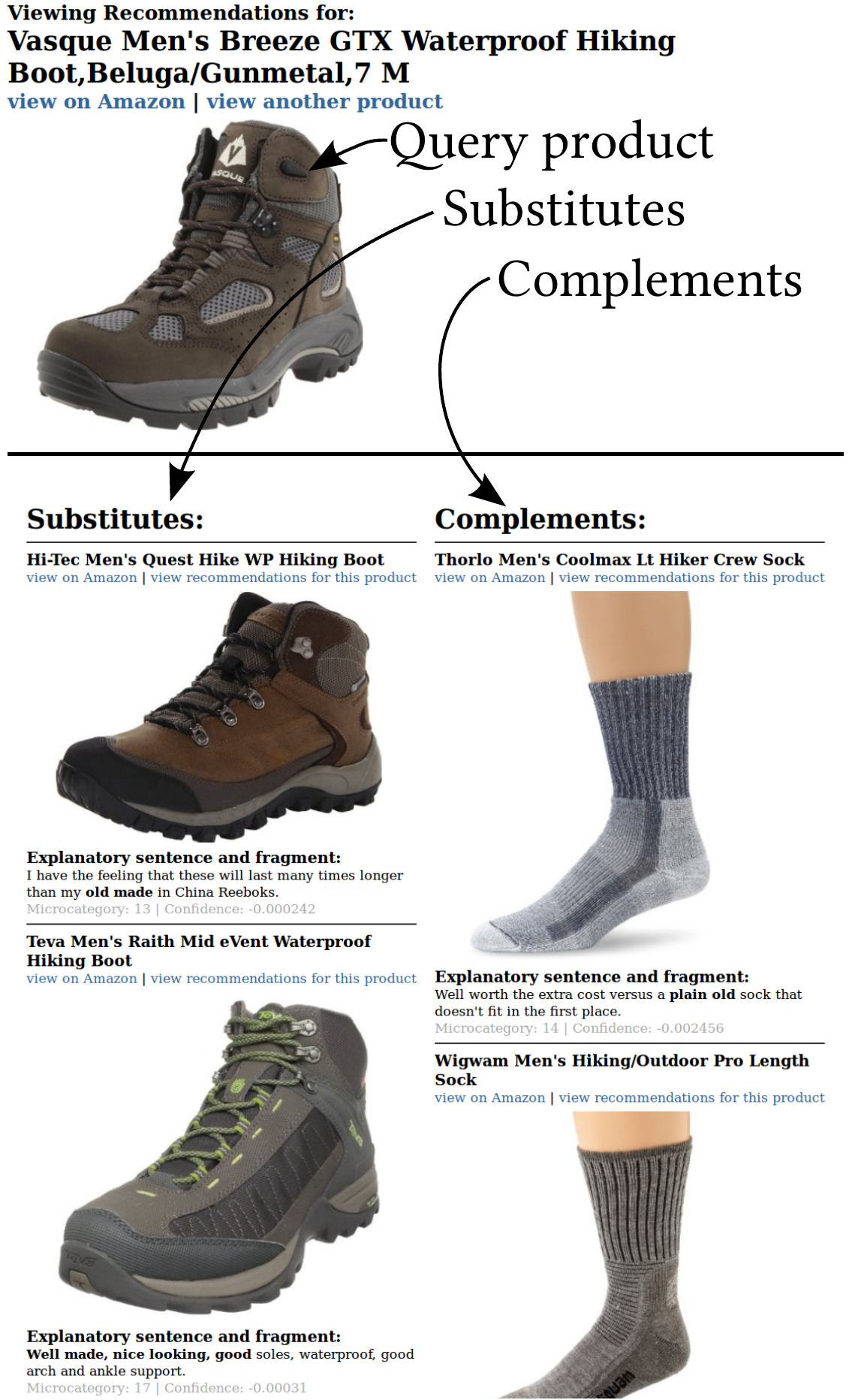}}}\vspace{2mm}
\caption{\modelname{} learns the concept of substitute and complement goods from product information (descriptions, reviews, etc.). Given a query item, \modelname{} allows us to generate substitute and complementary recommendations as shown above.
\label{fig:1}}

\end{figure}

An example of the output of \modelname{} is shown in Figure~\ref{fig:1}. Here, given a query item (a hiking boot), our system identifies a ranked list of potential substitutes (other hiking boots), and complements (heavy-duty socks, shoe polish, etc.).

We train \modelname{} on a large corpus of 9 million products from \emph{Amazon}, with 237 million connections derived from browsing and co-purchasing data.
We evaluate \modelname{} in terms of its accuracy at link prediction and ranking, where we find it to be significantly more accurate than alternatives.
We also use \modelname{} to build a product graph, 
where for every product we recommend a list of the most related complementary and substitutable products.
Finally, we show that \modelname{} can be applied in `cold-start' settings, by using other sources of text when reviews are unavailable.
Overall, we find that the most useful source of information to identify substitutes and complements is the text associated with each product (\ie, reviews, descriptions, and specifications), from which we are able to uncover the key features and relationships between products, and also to explain these relationships through textual signals. 

We envision several applications of the product graphs produced by our system. Our system can help users to navigate, explore and discover new and previously unknown products. Or, it can be used to identify interesting product combinations, \eg~we can recommend outfits by matching a shirt with complementary trousers and a jacket. And, our system can be used as a candidate-generation step in providing better and more context-relevant recommendations.

\section{Related work}

The basic task of a recommender system is to suggest relevant items to users, based on their opinions, context, and behavior. One component of this task is that of estimating users' ratings or rankings of products \cite{korenSurvey}, \eg~by matrix factorization \cite{Koren09} or collaborative filtering \cite{Linden03}. Our goal here is related but complementary to rating estimation as we aim to discover relations between products.

In principle the types of relationships in which we are interested can be mined from behavioral data, such as browsing and co-purchasing logs.
For example, \emph{Amazon} allows users to navigate between products through links such as `users who bought X also bought Y' and `users who viewed X also viewed Y'~\cite{Linden03}.
Such a `co-counting' solution, while simple, has a few shortcomings, for example it may produce noisy recommendations for infrequently-purchased products, and has limited ability to explain the recommendations it provides.
More sophisticated solutions have been proposed that make use of browsing and co-purchasing data (\eg~\cite{ebay}), but in contrast to such `behavioral-based' solutions our goal is to \emph{learn} the semantics of `what makes products related?' in order to generate new links, adapt to different notions of relatedness, and to understand and explain the features that cause humans to consider products to be related.

\emph{Topic models} are a fundamental building block of text modeling \cite{blei_lda,blei2007,blei2003hierarchical} and form the cornerstone of our model. A variety of works have used topic models within recommender systems (\eg~\cite{ester12c,ganu,brody,titov,titov08b,Gamon05,popescu05,recsysJulian}), though generally with the goal of predicting user \emph{ratings} (or opinions) rather than learning relationships between products as we do here.
More specifically, our work builds on topic models for networks: Block-LDA \cite{blockLDA}, topic-link LDA \cite{liu}, and relational topic models \cite{chang2009relational} all attempt to identify topics that explain the links in document networks. A promising line of work uses such ideas to model social and citation networks \cite{wang11,nubbi,dvu}. However, these methods have trouble scaling to large networks, while \modelname{} scales to corpora with millions of documents (products) and hundreds of millions of links.

Last, a related branch of work aims to enhance e-commerce using browsing data. For example, \cite{jagabat} aims to forecast commercial intent based on query logs; and in \cite{panigrahi} the authors use query data to identify attributes that are important to users in order to surface recommendations. While different in terms of the data and problem setting, such works are similar in that they uncover relationships from large sources of weakly-structured data.

\section{The Sceptre Model}

In the following we build \modelname{} gradually, but in such a way that at each step we are specifying a usable model. We highlight the differences between successive iterations of our model \textcolor{blue}{in blue}. We do this to emphasize the fact that while \modelname{} makes use of several interacting components, each of these components
brings an additional modeling aspect into the framework. Table \ref{tab:notation} describes the notation we use throughout the paper.

\subsection{High-level Overview}

We first present a high-level, conceptual view of \modelname{}, to explain the intuition behind the model before we fully specify it.

\xhdr{Topic Models} We use \emph{topic models} \cite{blei_lda} to discover topics from product reviews and other sources of text. Conceptually, this means that the text from millions of products can be clustered into a small number of dimensions, so that each product $i$ (and its text) can be represented by a topic vector $\theta_i$ encoding the extent to which reviews/descriptions of a given product discuss each of the topics.

\xhdr{Link Prediction} 
Topic models allow us to represent each product $i$ by a vector $\theta_i$. On top of this we can have a statistical model to predict properties about products. 
In our case, we use logistic regression to make predictions about \emph{pairs} of items, using features that combine the topics of two products $\theta_i$ and $\theta_j$ simultaneously. The classification task we are interested in is: \emph{does a relationship exist between $i$ and $j$?} Using \emph{pairwise} features of the products, \eg~$\psi(i,j) = \theta_j - \theta_i$, we build logistic classifiers such that $\inner{\beta}{\psi(i,j)}$ takes a positive value if $i$ and $j$ are connected by an edge. We further develop this model so that predicting the \emph{presence} of an edge and the \emph{direction} of an edge are treated as two separate tasks, to account for asymmetries and to help with interpretability.

Importantly, it should be noted that we do not train topic models and \emph{then} perform link prediction, but rather we define a joint objective such that we discover topics that are informative for our link prediction task. In this way our model uncovers topics that are good at `explaining' the relationships between products.

\xhdr{Micro-Categories}
An additional goal of \modelname{} is to be able to discover micro-categories of closely related products. We achieve this by using sparse representations of very high dimensional topic vectors for each product. We make use of explicit product hierarchies (such as the category tree available from \emph{Amazon}), where each node of the hierarchy has a small number of topics associated with it.
The hierarchical nature of the category tree means that topics associated with top-level nodes are general and broad, while topics associated with leaf categories focus on differentiating between subtle product features, which can be interpreted as micro-categories (\eg~different styles of running shoes).

\xhdr{Product graph}
Finally, we have a supervised learning framework to predict relationships between products. Discovering substitutes and complements then depends on the choices of graph we use to train the model, for which we collect several graphs of related products from \emph{Amazon}. For example, a co-purchasing graph such as `users frequently purchased $a$ and $b$ together' encodes some notion of complements, whereas a graph such as `users who viewed $a$ eventually purchased $b$' captures the notion of substitutes. Thus, for every product, we predict a list of complementary and substitutable products and collect them into a giant network of related products.

\subsection{Detailed Technical Description}

\subsubsection{Background: Latent Dirichlet Allocation}

Latent Dirichlet Allocation (LDA, \cite{blei_lda}) uncovers latent structure in document corpora. For the moment, `documents' shall be the set of reviews associated with a particular product. LDA associates each document in a corpus $d \in \mathcal T$ with a $K$-dimensional topic distribution $\theta_d$ (a stochastic vector, \ie, $\sum_k \theta_{d,k} = 1$), which encodes the fraction of words in $d$ that discuss each 
of the $K$ topics. That is, words in the document $d$ discuss topic $k$ with probability $\theta_{d,k}$.

Each topic $k$ also has an associated word distribution, $\phi_k$, which encodes the probability that a particular word is used for that topic. Finally, the topic distributions themselves ($\theta_d$) are assumed to be drawn from a Dirichlet prior.

The final model includes word distributions for each topic $\phi_k$, topic distributions for each document $\theta_d$, and topic assignments for each word $z_{d,j}$. Parameters $\Phi = \lbrace \theta, \phi \rbrace$ and topic assignments $z$ are traditionally updated via sampling \cite{blei_lda}. The likelihood of a particular text corpus $\mathcal T$ (given the word distribution $\phi$, topics $\theta$, and topic assignments for each word $z$) is then
\begin{equation}
 p(\mathcal T | \theta, \phi, z) = \prod_{d \in \mathcal T} \prod_{j = 1}^{N_d} \theta_{z_{d,j}}\cdot \phi_{z_{d,j}, w_{d,j}},
\label{eq:ldalikelihood}
\end{equation}
where we are multiplying over all documents in the corpus, and all words in each document. The two terms in the product are the likelihood of seeing these particular topics ($\theta_{z_{d,j}}$), and the likelihood of seeing these particular words for this topic ($\phi_{z_{d,j},w_{d,j}}$).

\begin{table}[t]
\begin{center}
\begin{tabular}{lp{0.75\linewidth}}
\toprule
Symbol & Description\\
\midrule
$d_i$ & document associated with an item (product) $i$\\
$\mathcal T$ & document corpus\\
$K$ & number of topics\\
$\theta_i$ & $K$-dimensional topic distribution for item $i$\\
$\phi_k$ & word distribution for topic $k$\\
$w_{d,j}$ & $j^{\text{th}}$ word of document $d$\\
$z_{d,j}$ & topic of the $j^{\text{th}}$ word document $d$\\
$N_d$ & number of words in document $d$\\
$F(x)$ & logistic (sigmoid) function, $1/(1 + e^{-x})$\\
$\mathcal E_g$ & observed edges in graph $g$\\
$\psi(i,j)$ & pairwise (undirected) features for items $i$ and $j$\\
$\varphi(i,j)$ & pairwise (directed) features for items $i$ and $j$\\
$\beta$ & logistic weights associated with $\psi(i,j)$\\
$\eta$ & logistic weights associated with $\varphi(i,j)$\\
\bottomrule
\end{tabular}
\end{center}
 \caption{Notation. \label{tab:notation}}
\end{table}

\subsubsection{Link Prediction with Topic Models}

`Supervised Topic Models' \cite{blei2007} allow topics to be discovered that are predictive of an output variable associated with each document. We propose a variant of a supervised topic model that identifies topics that are useful as features for link prediction. We choose an approach based on logistic regression because (1) It can be scaled to millions of documents/products by hundreds of millions of edges, and (2) It can be adapted to incorporate both latent features (topics) and manifest features (such as brand, price, and rating information), as well as arbitrary transformations and combinations of these features. Our goal here is to predict links, that is labels at the level of \emph{pairs} of products. In particular, we want to train logistic classifiers that for each pair of products $(i,j)$ predicts whether they are related ($y_{i,j} = 1$) or not ($y_{i,j} = 0$). For now we will consider the case where we are predicting just a single type of relationship and we will later generalize the model to predict multiple types of relationships (substitutes and complements) simultaneously. 

We want the topics associated with each product to be `useful' for logistic regression in the sense that we are able to learn a logistic regressor parametrized by $\beta$ that predicts $y_{i,j}$, using the topics $\theta_i$ and $\theta_j$ as features. That is, we want the logistic function
\begin{equation}
F_\beta(\theta_d) = \sigma(\inner{\beta}{\psi_\theta(i,j)})
\end{equation}
to match $y_{i,j}$ as closely as possible, where $\psi_\theta(i,j)$ is a pairwise feature vector describing the two products. We then aim to design features that encode the \emph{similarity} between the two products (documents). The specific choice we adopt is
\begin{equation}
\psi_\theta(i,j) = (1, \theta_{i,1}\cdot\theta_{j,1}, \theta_{i,2}\cdot\theta_{j,2}, \ldots, \theta_{i,K}\cdot\theta_{j,K}).
\label{eq:edge_feature}
\end{equation}
Intuitively, by defining our features to be the elementwise product between $\theta_i$ and $\theta_j$, we are saying that products with \emph{similar} topic vectors are likely to be linked. The logistic vector $\beta$ then determines \emph{which} topic memberships should should be similar (or dissimilar) in order for the products to be related.

Our goal then is to simultaneously optimize both topic distributions $\theta_d$ and logistic parameters $\beta$ to maximize the joint likelihood of topic memberships and relationships in the product graph:
\begin{multline}
L(y, \mathcal T | \beta, \theta, \phi, z) =  \overbrace{\prod_{d \in \mathcal T}\prod_{j = 1}^{N_d} \theta_{z_{d,j}}\phi_{z_{d,j}, w_{d,j}}   \vphantom{\prod_{(i,j) \in \mathcal E}\prod_{(i,j) \in \bar{\mathcal E}}}}^{\text{corpus likelihood}}\\
\underbrace{\prod_{\textcolor{black}{(i,j) \in \mathcal E}} F_\beta(\textcolor{black}{\psi_\theta(i,j)}) \prod_{\textcolor{black}{(i,j) \in \bar{\mathcal E}}} (1 - F_\beta(\textcolor{black}{\psi_\theta(i,j)}))}_{\text{logistic likelihood of the observed graph}}.
\label{eq:joint_objective_graph}
\end{multline}
This expression says that the review corpus should have high likelihood according to a topic model, but also that those topics should be useful as predictors in a logistic regressor that uses their similarity as features. In this way, we will intuitively discover topics that correspond to some concept of document `relatedness'.

This idea of jointly training topic and regression models is closely related to the model of \cite{recsysJulian}, where topics were discovered that are useful as parameters in a latent-factor recommender system. Roughly, in the model of \cite{recsysJulian}, a user would give a high rating to a product if their latent user parameters were similar to the topics discovered from reviews of that item; topics were then identified that were good at predicting users' ratings of items. The basic model of \eq{eq:joint_objective_graph} is similar in the sense that we are coupling parameters $\theta$ and $\beta$ in a joint likelihood in order to predict the output variable~$y$.

\xhdr{Directed vs. Undirected Graphs}
So far we have shown how to train topic models to predict links between products. However, the feature vector of \eq{eq:edge_feature} is symmetric ($\psi_\theta(i,j) = \psi_\theta(j,i)$), meaning that it is only useful for predicting \emph{undirected} relationships. However, none of the relationships we want to predict are necessarily symmetric. For example $y$ may be a good substitute for $x$ if $y$ is a similar product that is cheaper and better rated, but in this case $x$ would be a poor substitute for $y$. Or, while a replacement battery may be a good complement for a laptop, recommending a laptop to a user already purchasing a battery makes little sense. Thus we ought to account for such asymmetries in our model.

We model such asymmetries by first predicting whether two products are \emph{related}, and \emph{then} predicting in which direction the relation flows. That is, we predict
\begin{multline*}
p(a \text{\ has an edge toward\ } b) =\\
  p(a \text{\ is related to\ } b) \times p(\text{edge flows from\ } a \text{\ to\ } b \, | \,  a \text{\ is related to\ } b),
\end{multline*}
which we denote
\begin{equation}
p((a,b) \in \mathcal E) = \hspace{-1.5mm}\underbrace{p(a \leftrightarrow b)}_{\text{`are they related?'}} \hspace{-6.5mm} \overbrace{p(a \rightarrow b | a \leftrightarrow b),}^{\text{`does the edge flow in this direction?'}}\hspace{-11.5mm}
\end{equation}
where relations $(a,b) \in \mathcal E$ are now \emph{ordered} pairs (that may exist in both directions). We model relations in this way since
we expect different types of language or features to be useful for the two tasks---relatedness is a function of what two products have \emph{in common}, whereas the direction the link flows is a function of how the products \emph{differ}. 
Indeed, in practice we find that the second predictor $p(a \rightarrow b | a \leftrightarrow b)$ tends to pick up qualitative language that explains why one product is `better than' another, while the first tends to focus on high-level category specific topics.
Our objective now becomes
\begin{multline}
L(y, \mathcal T | \beta, \eta, \theta, \phi, z) =\\ \overbrace{\prod_{(i,j) \in \mathcal E} F^\leftrightarrow_\beta(\psi_\theta(i,j)) \textcolor{blue}{F^\rightarrow_\eta(\varphi_\theta(i,j)) (1 - F^\rightarrow_\eta(\varphi_\theta(j,i)))}}^{\text{positive relations ($F^\leftrightarrow$) and their direction of flow ($F^\rightarrow$)}}\\
\underbrace{\prod_{(i,j) \in \bar{\mathcal E}} (1 - F^\leftrightarrow_\beta(\psi_\theta(i,j)))}_{\text{non-relations}} \underbrace{\prod_{d \in \mathcal T}\prod_{j = 1}^{N_d} \theta_{z_{d,j}}\phi_{z_{d,j}, w_{d,j}}\vphantom{\prod_{(i,j) \in \bar{\mathcal E}}}}_{\text{corpus likelihood}}.
\label{eq:joint_objective_graph_directed}
\end{multline}
Here $F^\leftrightarrow$ is the same as in the previous section, though we have added $F^\rightarrow_\eta(\varphi_\theta(i,j))$ to predict edge directedness; this includes an additional logistic parameter vector $\eta$, as well as an additional feature vector $\varphi_\theta(i,j)$. The specific feature vector we use is
\begin{equation}
\varphi_\theta(i,j) = (1, \theta_{j,1} - \theta_{i,1}, \ldots, \theta_{j,K} - \theta_{i,K}),
\end{equation}
\ie~the direction in which an edge flows between two items is a function of the \emph{difference} between their topic vectors.

\xhdr{Incorporating Other Types of Features}
We can easily incorporate manifest features into our logistic regressors, which simply become additional dimensions in $\varphi_\theta(i,j)$. We include the difference in price, difference in average (star-) rating, and an indicator that takes the value 1 if the items were manufactured by different companies, allowing the model to capture the fact that users may navigate towards (or away from) cheaper products, better rated products, or products from a certain brand.

Our entire model ultimately captures the following simple intuition: (1) Users navigate between \emph{related} products, which should have similar topics (``what do $a$ and $b$ have in common?''), and (2) The \emph{direction} in which users navigate should be related to the difference between topics (``what does $b$ have that $a$ doesn't?''). Ultimately, all of the above machinery has been designed to discover topics and predictors that capture this intuition.

\xhdr{Learning Multiple Graphs}
Next we must generalize our approach to simultaneously learn multiple types of relationships. In our case we wish to discover a graph of products that users might purchase \emph{instead} (substitute products), as well as a graph of products users might purchase \emph{in addition} (complementary products).
Naturally, one could train models independently for each type of relationship. But then one would have two sets of topics, and two predictors that could be used to predict links in each graph.

Instead we decide to extend the model from the previous section so that it can predict multiple types of relations simultaneously. We do this by discovering a single set of topics that work well with \emph{multiple} logistic predictors. This is a small change from the previous model of \eq{eq:joint_objective_graph_directed}:
\begin{multline}
L(y, \mathcal T | \beta, \eta, \theta, \phi, z) = \overbrace{\prod_{d \in \mathcal T}\prod_{j = 1}^{N_d} \theta_{z_{d,j}}\phi_{z_{d,j}, w_{d,j}}}^{\text{corpus likelihood}}\\
 \textcolor{blue}{\prod_{g \in G}} \biggl\lbrace \prod_{(i,j) \in \mathcal E_{\textcolor{blue}{g}}} F^\leftrightarrow_{\beta_{\textcolor{blue}{g}}}(\psi_\theta(i,j)) F^\rightarrow_{\eta_{\textcolor{blue}{g}}}(\varphi_\theta(i,j)) (1 - F^\rightarrow_{\eta_{\textcolor{blue}{g}}}(\varphi_\theta(i,j)))\\
\hspace{8.5mm}\underbrace{\hspace{16mm}\prod_{(i,j) \in \bar{\mathcal E}_{\textcolor{blue}{g}}} (1 - F^\leftrightarrow_{\beta_{\textcolor{blue}{g}}}(\psi_\theta(i,j))) \biggr\rbrace.\hspace{16mm}}_{\text{accuracy of the predictors $\beta_g$ and $\eta_g$ \emph{for the graph} $g$}}\hspace{-0.5mm}
\label{eq:joint_objective_graph_directed_multigraph}
\end{multline}
where each graph $g\in G$ contains all relations of a particular type.

Note that we learn separate predictors $\beta_g$ and $\eta_g$ for each graph $g$, but we learn a \emph{single} set of topics ($\theta$) and features ($\psi$ and $\varphi$) that work well for all graphs simultaneously. We adopt this approach because it provides a larger training set that is more robust to overfitting compared to training two models separately. Moreover it means that both logistic regressors operate in the same feature space; this means that by carefully constructing our labeled training set (to be described in the following section), we can train the model not only to predict substitute and complementary relationships, but also to \emph{differentiate} between the two, by training it to identify substitutes as non-complements and vice versa.

\begin{figure}
\begin{center}
\includegraphics{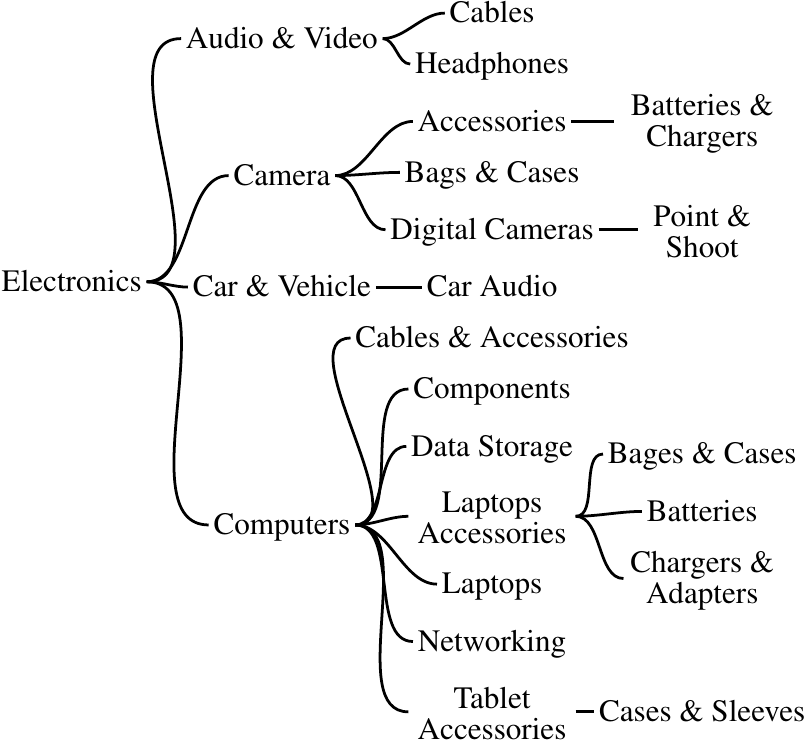}
\end{center}
\normalsize
\caption{Part of the product hierarchy for \emph{Amazon} Electronics products (the complete tree, even for Electronics alone, is too large to include here). \label{fig:hierarchy}}
\end{figure}

\begin{figure*}
\begin{center}
\includegraphics[width=0.95\textwidth]{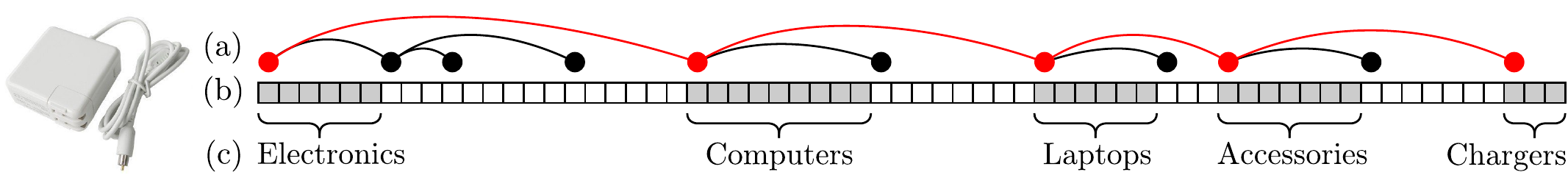}
\end{center}
 \caption{A demonstration of our topic hierarchy. A product (left) is shown with its associated topics (right). (a) the category tree (b) the topic vector (c) the product's ground-truth category. The product's position in the category tree is highlighted in red, and the set of topics that are `activated' is highlighted in gray. \label{fig:hierarchy_demo}}
\end{figure*}

\subsubsection{Sparse Topics via Product Hierarchies}
\label{sec:hierarchy}

Our goal is to learn topic models on corpora with millions of products and hundreds of topics. However, training models with hundreds of topics per product is not practical, nor is it realistic from a modeling perspective. Rather, each product should draw from a small number of topics, which can be encoded using a sparse representation \cite{bk:Supercharging}. To achieve this we enforce sparsity through a specific type of hierarchical topic representation that makes use of an explicit category tree, such as the one available from \emph{Amazon}.

An example of the product hierarchy we obtain from \emph{Amazon} is shown in Figure \ref{fig:hierarchy}. The category of each product is a node in this tree (though not necessarily a leaf node); some products may also belong to multiple categories simultaneously.

We build our topic representation using the following scheme: First, each product is represented by a path, or more simply a set of nodes, in the category tree. For products belonging to multiple categories, we take the union of those paths. Second, each topic is associated with a particular node in the category tree. Every time we observe, say, a thousand instances of a node, we associate an additional topic with that node, up to some maximum. In this way we have many topics associated with popular categories (like `Electronics') and fewer topics associated with more obscure categories. Third, we use a sparse representation for each product's topic vector. Specifically, if a product occupies a particular path in the category tree, then it can only draw words from topics associated with nodes in that path. In this way, even though our model may have hundreds of topics, only around 10-20 of these will be `active' for any particular product. This is not only necessary from a scalability standpoint, but it also helps the model quickly converge to meaningful topic representations.

This process is depicted in Figure \ref{fig:hierarchy_demo}. Here a product like a laptop charger draws from `generic' topics that apply to all electronics products, as well as topics that are specific to laptops or chargers; but it cannot draw from topics that are specific to mobile phones or laptop cases (for example). Thus all products have some high-level categories in common, but are also assumed to use their own unique sub-category specific language. Then, at the lowest level, each leaf node in the category tree is associated with multiple topics; thus we might learn several `microcategories' of laptop chargers, \eg~for different laptop types, price points, or brands. We present some examples of the types of microcategory we discover in Section \ref{sec:microcategories}.

\subsection{Optimization and Training}

Optimizing an objective such as the one in \eq{eq:joint_objective_graph_directed_multigraph} is a difficult task, for instance it is certainly non-convex.\footnote{It is smooth, and multiple local minima can be found by permuting the order of topics and logistic parameters.} We solve it using the following EM-like procedure, in which we alternate between optimizing the model parameters $\Theta = (\beta, \eta, \theta, \phi)$ and topic assignments (latent variables) $z$:
\begin{eqnarray}
 \text{update\ } \Theta^{(t)} = \argmax_{\Theta} l(y,\mathcal T | \beta, \eta, \theta, \phi, z^{(t-1)}) \label{eq:opt_step1}\\
 \text{sample\ } z^{(t)}_{d,j} \text{\ with probability\ } p(z_{d,j}^{(t)} = k) = \theta_{d,k}\phi_{k,w_{d,j}}^{(t)} \label{eq:opt_step2},
\end{eqnarray}
where $l(y, \mathcal T | \beta, \eta, \theta, \phi, z)$ is the log-likelihood from \eq{eq:joint_objective_graph_directed_multigraph}. To generate initial values for $\Theta$ and $z$ we initialize continuous parameters and topics uniformly at random (continuous parameters are sampled from $[0,1)$).

In the first step \eq{eq:opt_step1}, topic assignments for each word ($z$) are fixed. We fit the remaining terms, $\beta$, $\eta$, $\theta$, and $\phi$, by gradient ascent. We use the Hybrid LBFGS solver of \cite{hlbfgs}, a quasi-Newton method for non-linear optimization of problems with many variables \cite{lbfgs}. Computing the partial derivatives themselves, while computationally expensive, is na\"ively parallelizable over edges in $\mathcal E$ and documents (\ie, products) in $\mathcal T$.

The second step iterates through all products $d$ and all word positions $j$ and updates topic assignments. As with LDA, we assign each word to a topic (an integer between 1 and $K$) randomly, with probability proportional to the likelihood of that topic occurring with that word. The expression $\theta_{d,k}\phi_{k,w_{d,j}}$ is the probability of the topic $k$ for the product $d$ ($\theta_{d,k}$), multiplied by the probability of the word at position $j$ ($w_{d,j})$ being used in topic $k$ ($\phi_{k,w_{d,j}}$).

\section{Experiments}

Next we evaluate \modelname{}. We first describe the data as well as the baselines and then proceed with experimental results.

\subsection{Data}
\label{sec:data}

We use data crawled from \emph{Amazon.com}, whose characteristics are shown in Table \ref{tab:data}. This data was collected by performing a breadth-first search on the user-product-review graph until termination, meaning that it is a fairly comprehensive collection of English-language product data. We split the full dataset into top-level categories, \eg~Books, Movies, Music. We do this mainly for practical reasons, as it allows each model and dataset to fit in memory on a single machine (requiring around 64GB RAM and 2-3 days to run our largest experiment). Note that splitting the data in this way has little impact on performance, as there are few links that cross top-level categories, and the hierarchical nature of our model means that few parameters are shared across categories.

To obtain ground-truth for pairs of substitutable and complementary products we also crawl graphs of four types from \emph{Amazon}:
\begin{enumerate}
 \item `Users who viewed $x$ also viewed $y$'; 91M edges.
 \item `Users who viewed $x$ eventually bought $y$'; 8.18M edges.
 \item `Users who bought $x$ also bought $y$'; 133M edges.
 \item `Users frequently bought $x$ and $y$ together'; 4.6M edges.
\end{enumerate}
We refer to edges of type 1 and 2 as \emph{substitutes} and edges of type 3 or 4 as \emph{complements}, though we focus on `also viewed' and `also bought' links in our experiments, since these form the vast majority of the dataset. Note the minor differences between certain edge types, \eg~edges of type 4 indicate that two items were purchased as part of a single basket, rather than across sessions.

\subsection{Experimental Setting}

We split our training data ($\mathcal E$ and $\bar{\mathcal E}$) into 80\% training, 10\% validation, 10\% test, discarding products with fewer than 20 reviews. In all cases we report the error on the test set. The iterative fitting process described in (eqs.~\ref{eq:opt_step1} and \ref{eq:opt_step2}) continues until no further improvement is gained on the validation set.

\xhdr{Sampling Non-Edges} Since it is impractical to train on \emph{all} pairs of non-links, we start by building a balanced dataset by sampling as many non-links as there are links (\ie, $|\mathcal E| = |\bar{\mathcal E}|$).

However, we must be careful about how non-links (\ie, negative examples) are sampled. Sampling random pairs of unrelated products makes negative examples very `easy' to classify; rather, since we want the model to be able to \emph{differentiate} between edge types, we treat substitute links as negative examples of complementary edges and vice versa.
Thus, we explicitly train the model to identify substitutes as non-complements and vice versa (in addition to a random sample of non-edges). This does not make prediction `easier', but it helps the model to learn a better separation between the two edge types, by explicitly training it to learn distinct notions of the two concepts.

\begin{table}
\begin{center}
\begin{tabular}{lrrrr}
\toprule
Category         & Users  & Items & Reviews & Edges\\
\midrule
Men's Clothing   & 1.25M  & 371K  & 8.20M   & 8.22M\\
Women's Clothing & 1.82M  & 838K  & 14.5M   & 17.5M\\
Music            & 1.13M  & 557K  & 6.40M   & 7.98M\\
Movies           & 2.11M  & 208K  & 6.17M   & 4.96M\\
Electronics      & 4.25M  & 498K  & 11.4M   & 7.87M\\
Books            & 8.20M  & 2.37M & 25.9M   & 50.0M\\
\midrule
All              & 21.0M  & 9.35M & 144M    & 237M\\
\bottomrule
\end{tabular}
\end{center}

 \caption{Dataset statistics for a selection of categories on \emph{Amazon}. \label{tab:data}}
\end{table}

In the following, we consider both link prediction and ranking tasks: (1) to estimate for a pair of products whether they are related, and (2) for a given query, rank those items that are most likely to be related. We first describe the baselines we compare against.

\subsection{Baselines}

\xhdr{Random} Link probabilities $F_\beta$ and $F_\eta$ are replaced with random numbers between 0 and 1. Note that since both predictors have to `fire' to predict a relation, random classification identifies 75\% of directed edges as non-links; imbalance in the number of positive vs.~negative relations of each type (due to our sampling procedure described above) explains why the performance of random classification is slightly different across experiments.

\xhdr{LDA + logistic regression (LDA)} Rather than training topic models and logistic parameters simultaneously, this baseline \emph{first} trains a topic model and \emph{then} trains logistic classifiers on the pre-trained topics. This baseline assesses our claim that \modelname{} learns topics that are `good as' features for edge prediction, by comparing it to a model whose topics were \emph{not} trained specifically for this purpose. We used \emph{Vowpal Wabbit} to pre-train the topic model, and fit models with $K=100$ topics for each \emph{Amazon} category.

We also experimented with a baseline in which features were defined over \emph{words} rather than topics. That is, topics $\theta_i$ for each product are replaced by TF-IDF scores for words in its reviews \cite{iir}. Logistic parameters $\beta$ and $\eta$ are then trained to determine which tf-idf-weighted words are good at predicting the presence or absence of edges. This baseline was uniformly weaker than our other baselines, so we shall not discuss its performance further. 

\xhdr{Category-Tree (CT)} Since we make use of \emph{Amazon}'s category tree when building our model, it is worth questioning the extent to which the performance of our model owes to our decision to use this source of data. For this baseline, we compute the co-counts between categories $c_1 \rightarrow c_2$ that are observed in our training data. Then we predict that an edge exists if it is among the 50$^{\text{th}}$ percentile of most commonly co-occurring categories. In other words this baseline `lifts' links to the level of categories rather than individual products.\footnote{We experimented with several variations on this theme, and this approach yielded the best performance.}

\xhdr{Item-to-Item Collaborative Filtering (CF)} In 2003 \emph{Amazon} reported that their own recommendation solution was a collaborative-filtering approach, that identified items that had been browsed or purchased by similar sets of users \cite{Linden03}. This baseline follows the same procedure, though in lieu of actual browsing or purchasing data we consider sets of users who have \emph{reviewed} each item. We then proceed by computing for each pair of products $a$ and $b$ the cosine similarity between the set of users who reviewed $a$ and the set of users who reviewed $b$. Sorting by similarity generates a ranked list of recommendations for each product. Since this method is not probabilistic we only report its performance at ranking tasks.

\subsection{Link Prediction and Ranking}
\label{sec:predictionandranking}

\xhdr{Link Prediction} Our first goal is to predict for a given pair of products $(a,b)$, and a graph type $g$, whether there is a link from $a$ to $b$ in $\mathcal E_g$. We optimize exactly the objective in \eq{eq:joint_objective_graph_directed_multigraph}. Note that a prediction is correct when
\begin{itemize}
\item for each test edge (in each graph): $a \rightarrow b$,\\ $F^\leftrightarrow_\theta(\psi(a,b),\beta) > 0$ \emph{\textbf{and}} $F^\rightarrow_\theta(\varphi(a,b),\eta) > 0$
\item for each \emph{non}-edge $a \not\rightarrow b$,\\ $F^\leftrightarrow_\theta(\psi(a,b),\beta) \leq 0$ \emph{\textbf{or}} $F^\rightarrow_\theta(\varphi(a,b),\eta) \leq 0$,
\end{itemize}
in other words the model must correctly predict both that the link exists and its direction.

Results are shown in Table \ref{tab:results_links} for each of the datasets in Table \ref{tab:data}. We also show results from `Baby' clothes, to demonstrate that performance does not degrade significantly on a (relatively) smaller dataset (43k products). `Substitute' links were unavailable for the vast majority of products from Music and Movies in our crawl, so results are not shown. We summarize the main findings from this table as follows:
\begin{enumerate}
 \item \modelname{} is able to accurately predict both substitute and complement links across all product categories, with performance being especially accurate for clothing and electronics products. Accuracy is between 85.57-96.76\% for the binary prediction tasks we considered.
 \item Prediction of `substitute' links is uniformly more accurate than `complement' links for all methods, both in absolute (left two columns) and relative (right two columns) terms. This matches our intuition that substitute links should be `easier' to predict, as they essentially correspond to some notion of similarity, whereas the semantics of complements are more subtle.
 \item The performance of the baselines is variable. For substitute links, our LDA baseline obtains reasonable performance on Books and Electronics, whereas the Category Tree (CT) baseline is better for Clothing. In fact, the CT baseline performs surprisingly well at predicting substitute links, for the simple reason that substitutable products often belong to the same category as each other.
 \item None of the baselines yield satisfactory performance when predicting complement links. Thus we conclude that neither the topics uncovered by a regular topic model, nor the category tree alone are capable of capturing the subtle notions of what makes items complementary.
\end{enumerate}
Ultimately we conclude that each of the parts of \modelname{} contribute to its accurate performance. Category information is helpful, but alone is not useful to predict complements; and simultaneous training of topic models and link prediction is necessary to learn useful topic representations.

\begin{table}[t]
\begin{center}
\renewcommand{\tabcolsep}{1.75mm}
\begin{tabular}{lr|rr|rr}
 &  & \multicolumn{2}{c|}{Accuracy} & \multicolumn{2}{c}{\parbox{0.25\linewidth}{\centering Error reduction vs.~random}}\\[2mm]
Category & Method & Subst. & Compl. & Subst. & Compl.\\
\midrule
\multirow{4}{10mm}{Men's Clothing}
 & Random & 60.27\% & 57.70\% & 0.0\% & 0.0\% \\
 & LDA & 70.62\% & 65.95\% & 26.05\% & 19.50\% \\
 & CT & 78.69\% & 61.06\% & 46.38\% & 7.946\% \\
 & \modelname{} & 96.69\% & 94.06\% & 91.67\% & 85.97\% \\[2mm]
\multirow{4}{10mm}{Women's Clothing}
 & Random & 60.35\% & 56.67\% & 0.0\% & 0.0\% \\
 & LDA & 70.70\% & 64.80\% & 26.11\% & 18.75\% \\
 & CT & 81.05\% & 69.08\% & 52.21\% & 28.63\% \\
 & \modelname{} & 95.87\% & 94.14\% & 89.59\% & 86.47\% \\[2mm]
\multirow{4}{10mm}{Music}
 & Random & \multicolumn{1}{c}{-} & 50.18\% & \multicolumn{1}{c}{-} & 0.0\% \\
 & LDA & \multicolumn{1}{c}{-} & 52.39\% & \multicolumn{1}{c}{-} & 4.428\% \\
 & CT & \multicolumn{1}{c}{-} & 57.02\% & \multicolumn{1}{c}{-} & 13.71\% \\
 & \modelname{} & \multicolumn{1}{c}{-} & 90.43\% & \multicolumn{1}{c}{-} & 80.78\% \\[2mm]
\multirow{4}{10mm}{Movies}
 & Random & \multicolumn{1}{c}{-} & 51.22\% & \multicolumn{1}{c}{-} & 0.0\% \\
 & LDA & \multicolumn{1}{c}{-} & 54.26\% & \multicolumn{1}{c}{-} & 6.235\% \\
 & CT & \multicolumn{1}{c}{-} & 66.34\% & \multicolumn{1}{c}{-} & 30.99\% \\
 & \modelname{} & \multicolumn{1}{c}{-} & 85.57\% & \multicolumn{1}{c}{-} & 70.42\% \\[2mm]
\multirow{4}{10mm}{Electronics}
 & Random & 69.98\% & 55.67\% & 0.0\% & 0.0\% \\
 & LDA & 89.90\% & 61.90\% & 66.35\% & 14.06\% \\
 & CT & 87.26\% & 60.18\% & 57.57\% & 10.17\% \\
 & \modelname{} & 95.70\% & 88.80\% & 85.69\% & 74.74\% \\[2mm]
\multirow{4}{10mm}{Books}
 & Random & 69.93\% & 55.35\% & 0.0\% & 0.0\% \\
 & LDA & 89.91\% & 60.59\% & 66.47\% & 11.75\% \\
 & CT & 87.80\% & 66.28\% & 59.42\% & 24.49\% \\
 & \modelname{} & 93.76\% & 89.86\% & 79.25\% & 77.29\% \\[2mm]
\multirow{4}{10mm}{Baby Clothes}
 & random & 62.93\% & 52.47\% & 0.0\% & 0.0\% \\
 & LDA & 75.86\% & 54.73\% & 34.89\% & 4.75\% \\
 & CT & 79.31\% & 64.56\% & 44.18\% & 25.43\% \\
 & \modelname{} & 92.18\% & 93.65\% & 78.91\% & 86.65\% \\
 \midrule
Average & \modelname{} & 94.83\% & 90.23\% & 85.02\% & 80.33\%\\
\end{tabular}
\normalsize
\end{center}
 \caption{Link prediction accuracy for substitute and complement links (the former are not available for the majority of Music/Movies products in our dataset). Absolute performance is shown at left, reduction in error vs.~random classification at right. \label{tab:results_links}}
\end{table}

\begin{table}[t]
\begin{center}
\renewcommand{\tabcolsep}{1.75mm}
\begin{tabular}{l|rr|rr}
 &  \multicolumn{2}{c|}{Accuracy} & \multicolumn{2}{c}{\parbox{0.25\linewidth}{\centering Error reduction vs.~random}}\\[2mm]
Category & Subst. & Compl. & Subst. & Compl.\\
 \midrule
Electronics, cold-start & 91.28\% & 93.22\% & 70.95\% & 84.71\%\\
Books, cold-start & 96.76\% & 93.67\% & 89.22\% & 85.81\%\\
\end{tabular}
\normalsize
\end{center}
 \caption{Link prediction accuracy using cold-start data (manufacturer's and editorial descriptions). \label{tab:results_coldstart}}
\end{table}

\xhdr{Ranking}
\label{sec:ranking}
In many applications distinguishing links from non-links is not enough as for each product we would like to recommend a limited number of substitutes and complements. Thus, it is  important that relevant items (\ie, relevant relations) are ranked higher than irrelevant ones, regardless of the likelihood that the model assigns to each recommendation.

A standard measure to evaluate ranking methods is the precision@k. Given a set of recommended relations of a given type $\mathit{rec}$, and a set of known-relevant products $\mathit{rel}$ (\ie, ground-truth links) the \emph{precision} is defined as
\begin{equation}
 \text{precision} = |\mathit{rel} \cap \mathit{rec}| / |\mathit{rec}|,
\end{equation}
\ie, the fraction of recommended relations that were relevant. The precision@k is then the precision obtained given a fixed budget, \ie, when $|\mathit{rec}| = k$.
This is relevant when only a small number of recommendations can be surfaced to the user, where it is important that relevant products appear amongst the first few suggestions.

\begin{figure}[t]
 \includegraphics[width=0.49\linewidth]{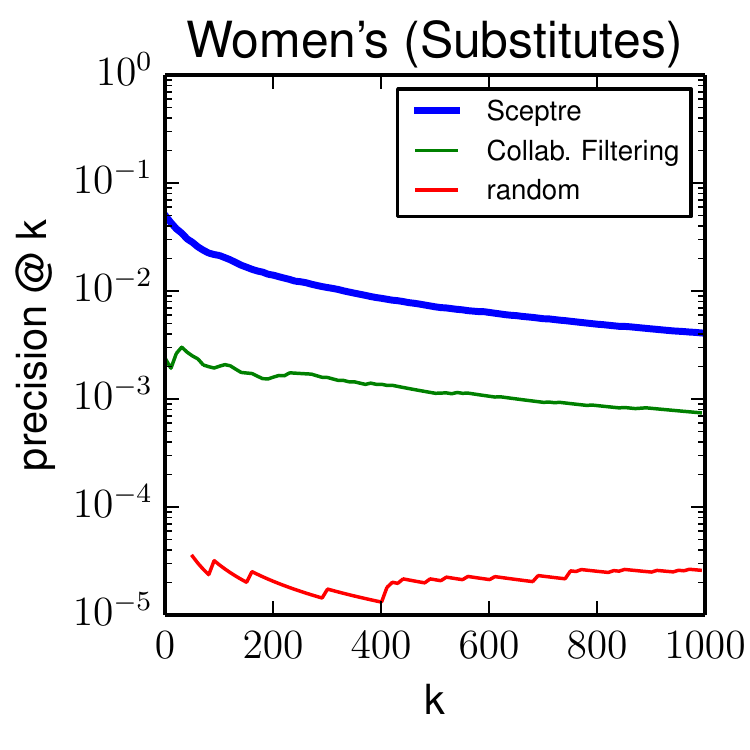}\includegraphics[width=0.49\linewidth]{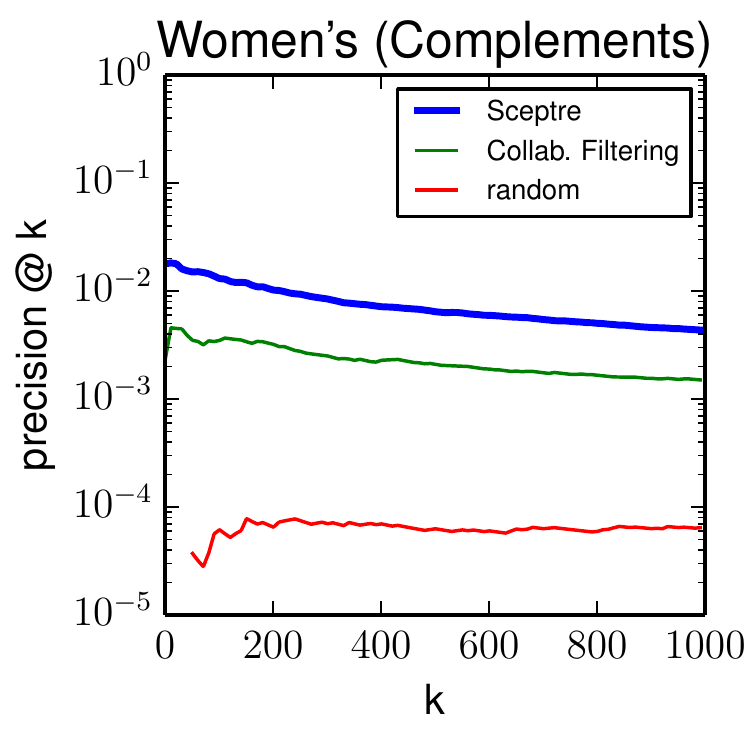}
 \includegraphics[width=0.49\linewidth]{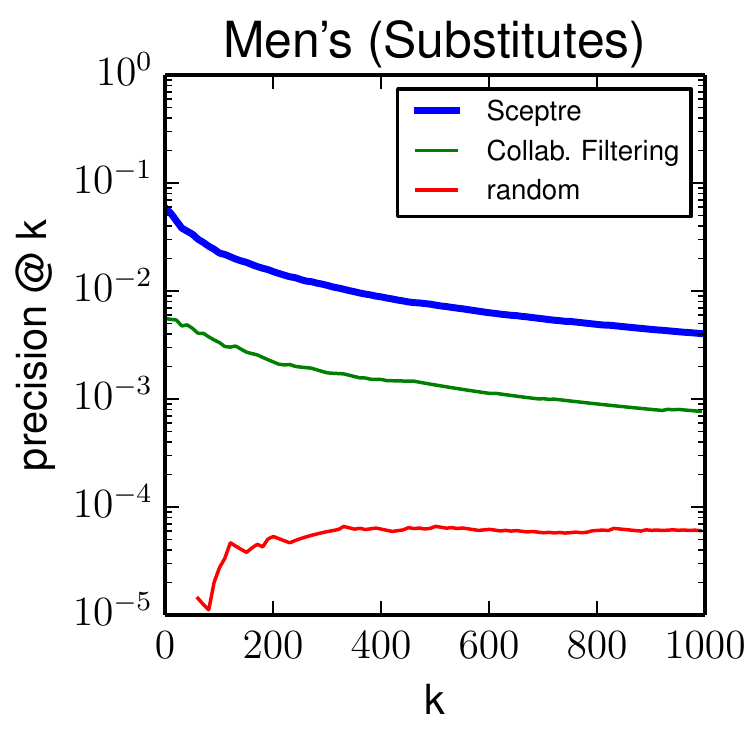}\includegraphics[width=0.49\linewidth]{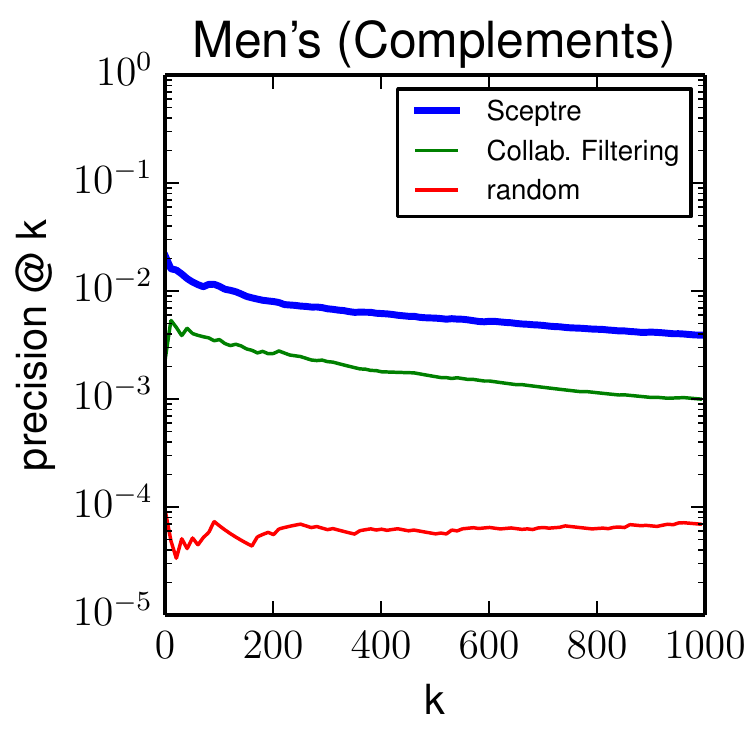}
 \caption{Precision@k for Women's and Men's clothing. \label{fig:prec}}
\end{figure}

Figure \ref{fig:prec} reports the precision@k on Men's and Women's clothing. Note that we naturally discard candidate links that appeared during training. This leaves only a small number of relevant products for each query item in the corpus---the random baseline (which up to noise should be flat) has precision around $5\times 10^{-5}$, indicating that only around 5 in 100,000 products are labeled as `relevant' in this experiment. This, in addition to the fact that many relevant items may not be labeled as such (there are presumably \emph{thousands} of pairs of substitutable pants in our corpus, but only 30 or so are recommended for each product) highlights the incredible difficulty of obtaining high precision scores for this task.

Overall, collaborative filtering is one to two orders-of-magnitude more accurate than random rankings, while \modelname{} is an order of magnitude more accurate again (our LDA and TF-IDF baselines were less accurate than collaborative filtering and are not shown).\footnote{Note that collaborative filtering is done here at the level of \emph{reviewed} products, which is naturally much sparser than the purchase and browsing data used to produce the ground-truth.}

Examples of recommendations generated by \modelname{} are shown in Figure \ref{fig:examples}.

\subsection{Cold-Start Prediction without Reviews}

Although it is promising that we are able to recommend substitutes and complements from the text of product reviews, an obvious question remains as to what can be done for \emph{new} products, that do not yet have any reviews associated with them, known as the `cold-start' problem in recommender systems~\cite{ester,levi12,park,schein,ke}. To address this problem, we note that \modelname{} merely requires that we have some source of text associated with each linked item in order to learn a model of which products are likely to be related.

To evaluate the possibility of using sources of text other than reviews, we collected descriptive text about each item in our \emph{Amazon Books} catalog, including blurbs, `about the author' snippets, and editorial reviews. 
We also collected manufacturer's descriptions for a subset of our Electronics data.
Training on these sources of data, \modelname{} obtained accuracies between 91.28\% and 93.67\% at predicting substitutes and complements (see Table \ref{tab:results_coldstart}).
This result implies that training is possible on diverse sources of text beyond just product reviews, and that \modelname{} can be applied in cold-start settings, even when no reviews are available.\footnote{Note that this is not directly comparable to review-based results since different subsets of our corpus have reviews vs.~descriptions.}

\begin{table*}[t]
\setlength{\tabcolsep}{1.3pt}
\begin{center}
Electronics
\small
  \begin{tabular}{ccccccccccc}
\toprule
e111 & e92 & e75 & e79 & e78 & e50 & e69 & e85 & e96 & e89 & e99\\
cameras  & portable speakers & cases & \parbox{15mm}{\centering Samsung cases} & \parbox{15mm}{\centering heavy-duty cases} & styli & batteries & portable radios & car radios & \parbox{14.3mm}{\centering high-end headphones} & \parbox{14.3mm}{\centering budget headphones}\\
\midrule
camera   & little speaker   & leather    & Galaxy         & Otterbox         & pen            & batteries    & radio           & radio      & bass                 & bass\\
zoom     & bose             & case       & elastic        & Defender         & tip            & battery      & weather         & Pioneer    & Sennheiser           & Skullcandy\\
pictures & portable speaker & soft       & magnets        & protection       & Bamboo         & charged      & crank           & factory    & Bose                 & sound\\
Kodak    & small speaker    & Roocase    & Samsung        & bulky            & Wacom          & rechargeable & solar           & Metra      & Shure                & bud\\
Canon    & sound            & velcro     & leather        & kids             & styli          & oem          & Eton            & Ford       & Beats                & outside noise\\
flash    & iHome            & closed     & closed         & shell            & gloves         & Sanyo        & Baofeng         & dash       & Koss                 & another pair\\
digital  & bass             & material   & auto           & Survivor         & \parbox{16mm}{\centering\ \hspace{-10mm}Friendly Swede\hspace{-10mm}\ } & Lenmar       & radio reception & Honda      & Akg                  & comfortable\\
optical  & wireless speaker & snug       & closing        & protected        & pencil         & alkaline     & miles           & Jeep       & music                & gym\\
taken    & great speaker    & protection & elastic strap  & safe             & capacitive     & Energizer    & fm              & wiring     & classical            & Beats\\
picture  & mini speaker     & standing   & cover          & protective       & precise        & full charge  & alert           & deck       & Klipsch              & head\\
\bottomrule 
 \end{tabular}
 \normalsize

\ \\
\setlength{\tabcolsep}{1.7pt}
 Men's clothing
 \small
 \begin{tabular}{ccccccccccc}
\toprule
c44 & c107 & c75 & c49 & c52 & c110 & c156 & c134 & c133 & c24 & c9\\
dress shirts & dress shoes & dress pants & \parbox{15mm}{\centering three-wolf\\ shirt}  & polo shirts    & boots       & \parbox{15mm}{\centering minimalist running} & \parbox{15mm}{\centering athletic running} & \parbox{15mm}{\centering sports shoes} & \parbox{12mm}{\centering generic\\ clothing} & \parbox{12mm}{\centering generic\\ clothing}\\
\midrule
sleeves     & leather      & expandable       & wolf         & Polo          & Bates       & running      & Balance        & court        & dry       & same\\
arms        & sole         & \parbox{11mm}{\centering\ \hspace{-13mm}expandable waist\hspace{-13mm}\ } & moon         & Lauren        & Red Wing    & trail        & New            & play         & cold        & durable\\
neck        & dress        & Dockers          & three        & Ralph         & leather     & barefoot     & wide           & Nike         & working         & store\\
shoulders   & brown        & iron             & power        & Beene         & good boot   & Altra        & running        & running shoe & short & different\\
dress shirt & dress shoe   & khaki            & trailer      & nice shirt    & casual boot & running shoe & series         & running      & hot             & two\\
dress       & polish       & stretch waist    & hair         & Geoffrey      & dress boot  & minimalist   & feet           & games        & weather          & brand\\
jacket      & brown pair   & hidden           & man          & great shirt   & right boot  & zero drop    & usa            & light shoe   & tight          & comfort\\
long sleeve & toe          & ironed           & short-sleeve & quality shirt & motorcycle  & road         & \parbox{13mm}{\centering\ \hspace{-5mm}cross training\hspace{-5mm}\ } & great shoe   & cool        & fine\\
iron        & looking shoe & dress pant       & magic        & white shirt   & Wings       & glove        & \parbox{13mm}{\centering\ \hspace{-5mm}athletic shoe\hspace{-5mm}\ } & support      & down          & tight\\
tucked      & formal       & elastic waist    & powerful     & fitted shirt  & Rangers     & run          & cross          & miles        & regular           & another pair\\
\bottomrule 
 \end{tabular}
 \end{center}
 \caption{A selection of topics from Electronics and Men's Clothing along with our labels for each topic. Top 10 words/bigrams from each topic are shown after subtracting the background distribution. Capital letters denote brand names (Bamboo, Wacom, Red Wing, etc.). \label{tab:topics}}
 \normalsize
\end{table*}

\subsection{Topic Analysis}
\label{sec:microcategories}
Next we analyze the types of topics discovered by \modelname{}. As we recall from Section \ref{sec:hierarchy}, each topic is associated with a node in \emph{Amazon}'s category tree. But, just as a high-level category such as clothing can naturally be separated into finer-grained categories like socks, shoes, hats, pants, dresses (etc.), we hope that \modelname{} will discover even subtler groupings of products
that are not immediately obvious from the hand-labeled category hierarchy.

Table \ref{tab:topics} shows some of the topics discovered by \modelname{}, on two \emph{Amazon} categories: Electronics and Men's Clothing. We pruned our dictionary by using adjectives, nouns, and adjective-noun pairs (as determined by WordNet \cite{WordNet}), as well as any words appearing in the `brand' field of our crawled data. For visualization we compute the 10 highest-weight words from all topics, after first subtracting a `background' topic containing the average weight across all topics. That is for each topic $k$ we report the 10 highest values of
\begin{equation}
 \phi_k - \hspace{-6mm}\underbrace{\frac{1}{K} \sum_{k'} \phi_{k'}.}_{\text{background word distribution}}\hspace{-6mm}
\end{equation}
By doing so, stopwords and other language common to all topics appears only in the background distribution.

The topics we obtain are closely aligned with categories from \emph{Amazon} (\eg~electronics topic e111, or clothing topic c110), though this is to be expected since our topic model is built on top of an explicit hierarchy as in Figure \ref{fig:hierarchy_demo}. However, we note that finer-grained `microcategories' are discovered that are not explicitly present in the hierarchy, \eg~high-end headphones are distinguished from che\-aper models (e89 and e99), and running shoes are distinguished based on distinct running styles (c133, c134, and c156).

We also note that brand words predominate in several topics, \eg~high-end headphones can be identified by words like `Sennhei\-ser', `AKG' etc. (e89), and high-end t-shirts can be identified by words like `Ralph Lauren' and `Geoffrey Beene' (c52). At the extreme end, a \emph{single product} may have its own topic, \eg~the popular `three-wolf moon' shirt (c49), whose reviews have already inspired academic discussion \cite{reyes11}. Here the product's high popularity and unique word distribution means that dedicating it an entire topic substantially increases the corpus likelihood in \eq{eq:joint_objective_graph_directed_multigraph}. Note that we only show a fraction of the full set of topics discovered by our model; other common camera brands (etc.) are covered among the large range of topics not shown here.

Finally, while some topics are highly specific, like those referring to individual products or brands, others are more generic, such as clothing topics c9 and c24. Such topics tend to appear toward the top of the category hierarchy (see Fig.~\ref{fig:hierarchy_demo}), for example the topic c9 is associated with the `Clothing' node, whereas c24 is associated with its child, `Clothing: Men', of which all other topics in Table \ref{tab:topics} are descendants. Intuitively, these are much like `background' distributions, containing words that are relevant to the majority of clothing products, like durability, fit, warmth, color, packaging, etc.

\subsection{User Study}

Finally we perform a user study to evaluate the quality of the recommendations produced by \modelname{}. Naturally we would not expect that a fully-supervised algorithm would produce predictions that were somehow `better' than the ground-truth used to train it. However, we hope \modelname{} may correct for some noise in the ground-truth, since while users may often buy multiple pairs of jeans together (for example) we are explicitly training the system to identify complementary items that would \emph{not} be substitutable.

\begin{figure}
\textcolor{white}{latex sucks}\hspace{2mm} \includegraphics[width=0.7\linewidth]{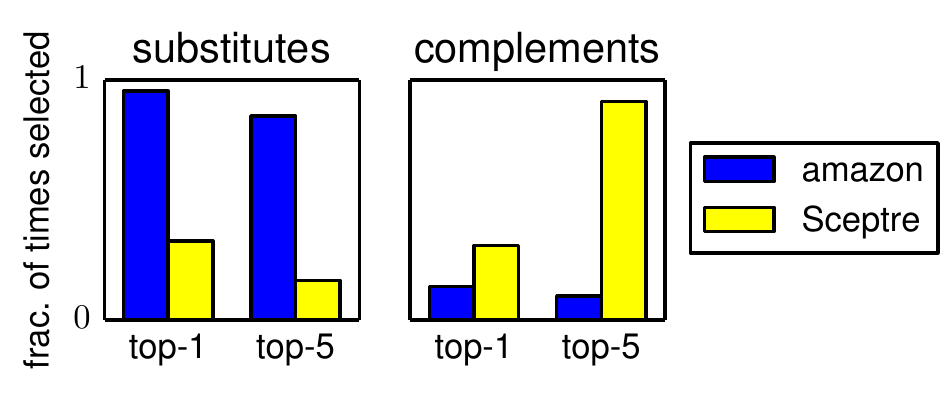}
\caption{Results of our user study. Users were asked to select which recommendations (ours or Amazon's) were preferable substitutes/complements (users could also select neither or both).\label{fig:mturk}}
\end{figure}

We used Mechanical Turk to compare \modelname{}'s recommendations to \emph{Amazon}'s `also viewed' and `also bought' suggestions, for a random sample of 200 Clothing items. Human judges identified which recommendations they felt were acceptable substitutes and complements (surfaced in a random order without labels; a screenshot is shown in Fig.~\ref{fig:examples}d). Judges evaluated the top recommendation, and top-5 recommendations separately, yielding results shown in Figure \ref{fig:mturk}.

We see here that Amazon's `also viewed' links generate preferable substitutes, indicating that large volumes of browsing data yield acceptable substitute relationships with relatively little noise. On the other hand, \modelname{}'s complement links are overwhelmingly preferred, suggesting that our decision to model complements as non-substitu\-tes qualitatively improves performance.

\section{Building the product graph}
\label{sec:interface}

\begin{figure*}[t]
\setlength{\fboxsep}{1mm}
        \centering
\subfloat[Men's clothing]
{
\framebox{\adjustbox{trim={.065\width} {.02\height} {.065\width} {.12\height},clip}{\includegraphics[width=0.202\textwidth]{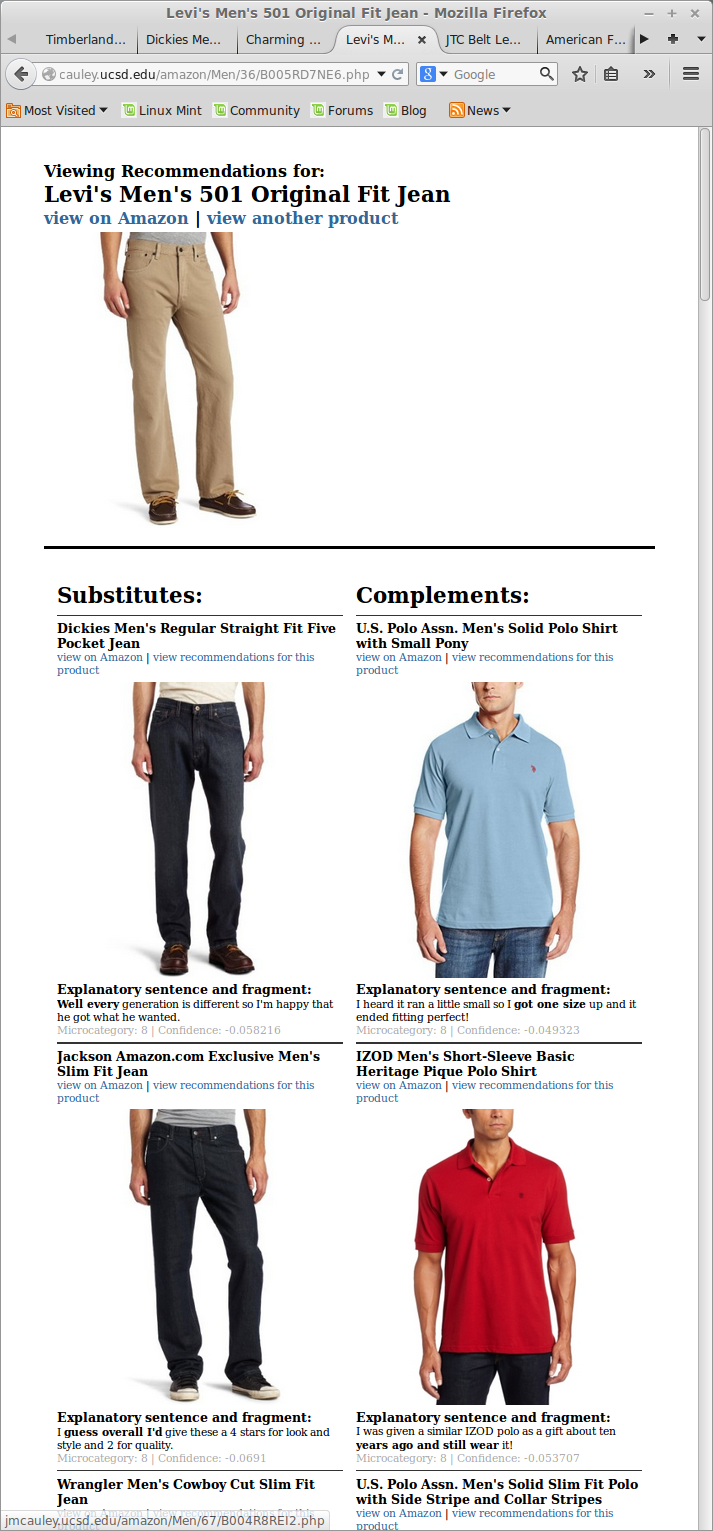}}}

}%
\subfloat[Women's clothing]
{

\framebox{\adjustbox{trim={.065\width} {.02\height} {.065\width} {.12\height},clip}{\includegraphics[width=0.202\textwidth]{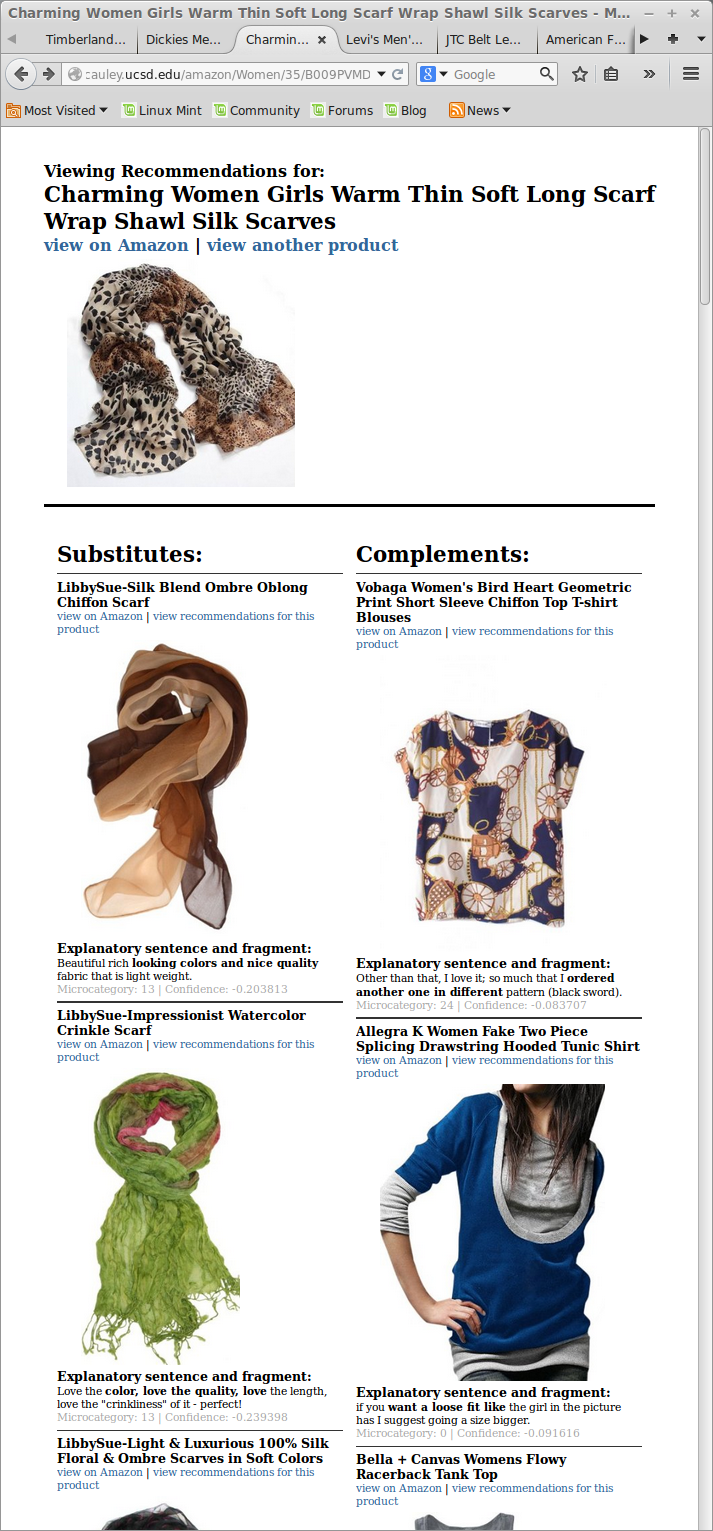}}}
}%
\subfloat[Electronics]
{
\framebox{\adjustbox{trim={.065\width} {.013\height} {.065\width} {.127\height},clip}{\includegraphics[width=0.202\textwidth]{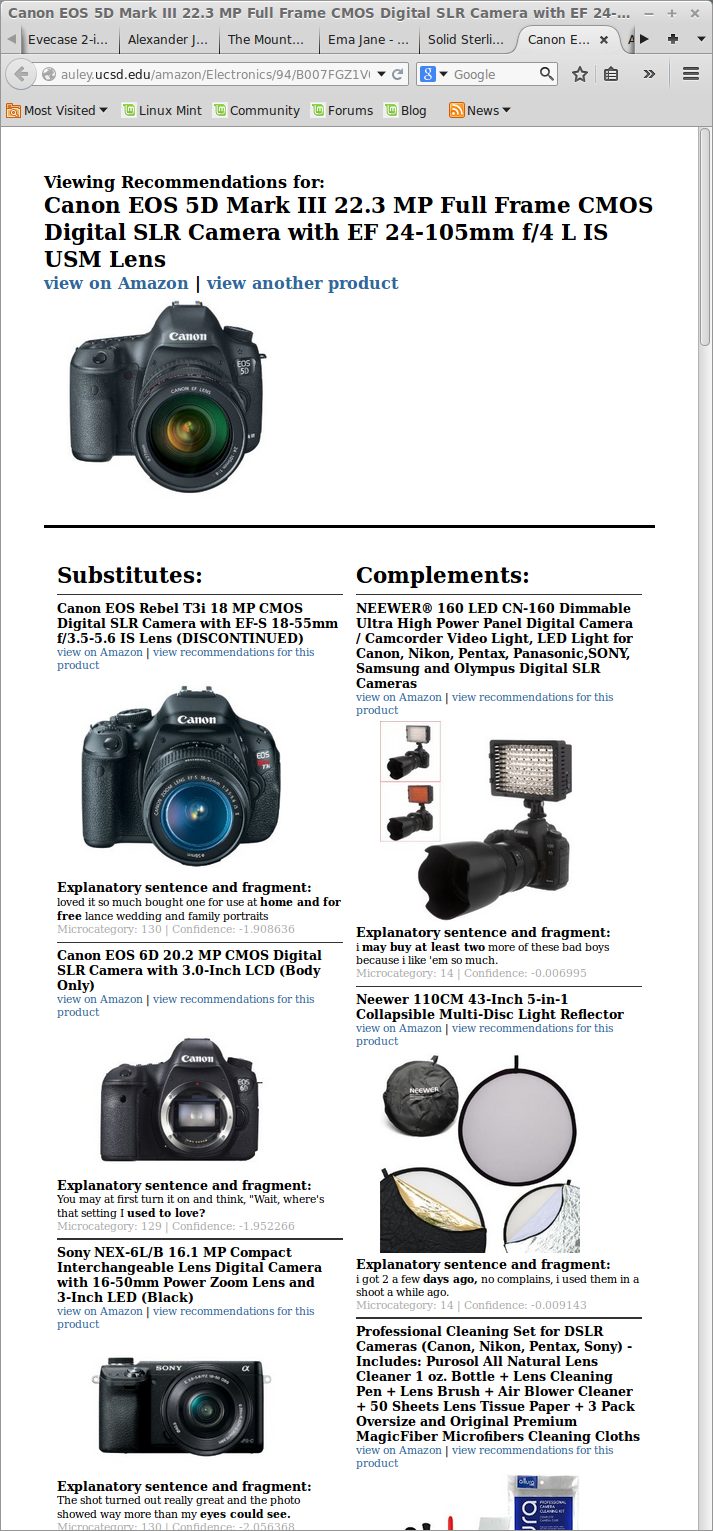}}}\hspace{0.8mm}
\framebox{\adjustbox{trim={.065\width} {.013\height} {.065\width} {.127\height},clip}{\includegraphics[width=0.202\textwidth]{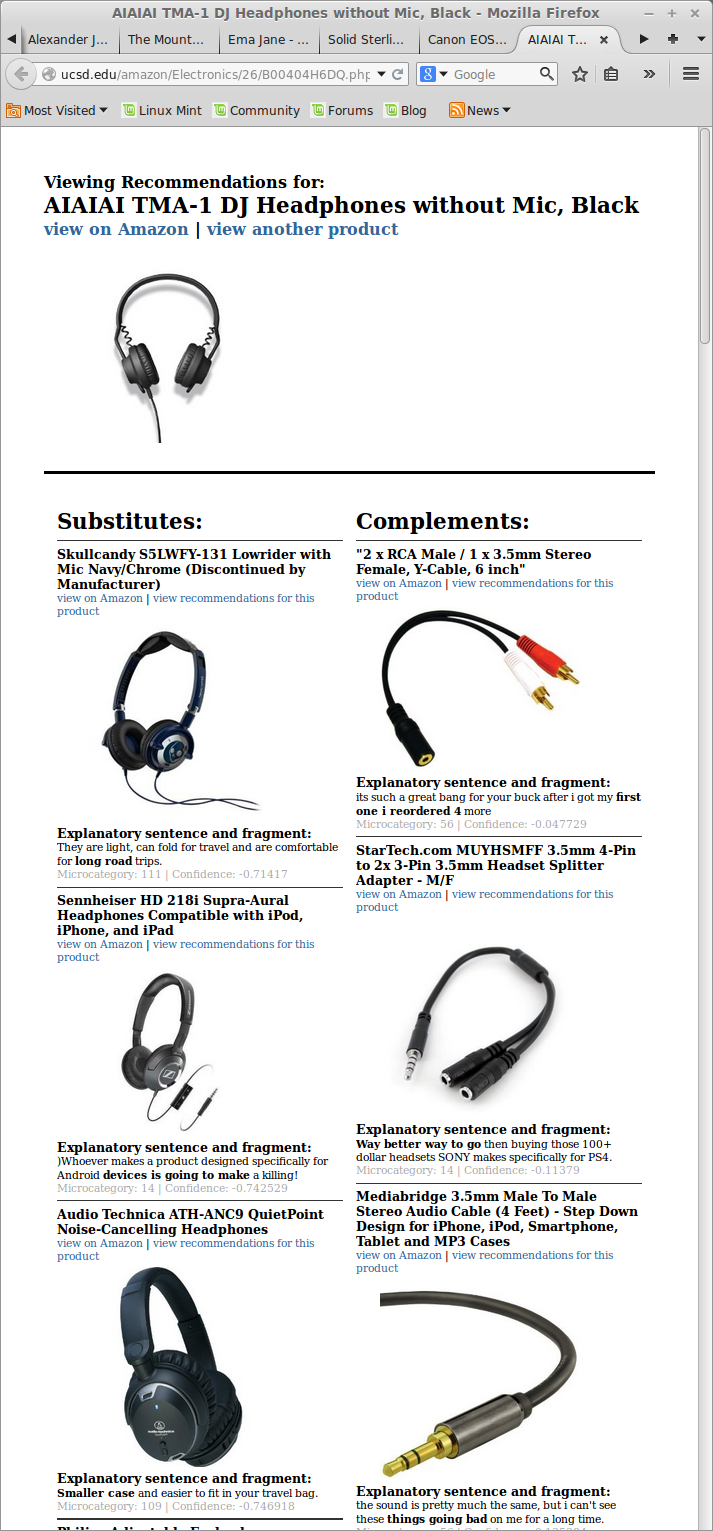}}}
}
\subfloat[mturk interface]
{
\framebox{\adjustbox{trim={.02\width} {.00\height} {.1\width} {.00\height},clip}{\includegraphics[width=0.2132\textwidth]{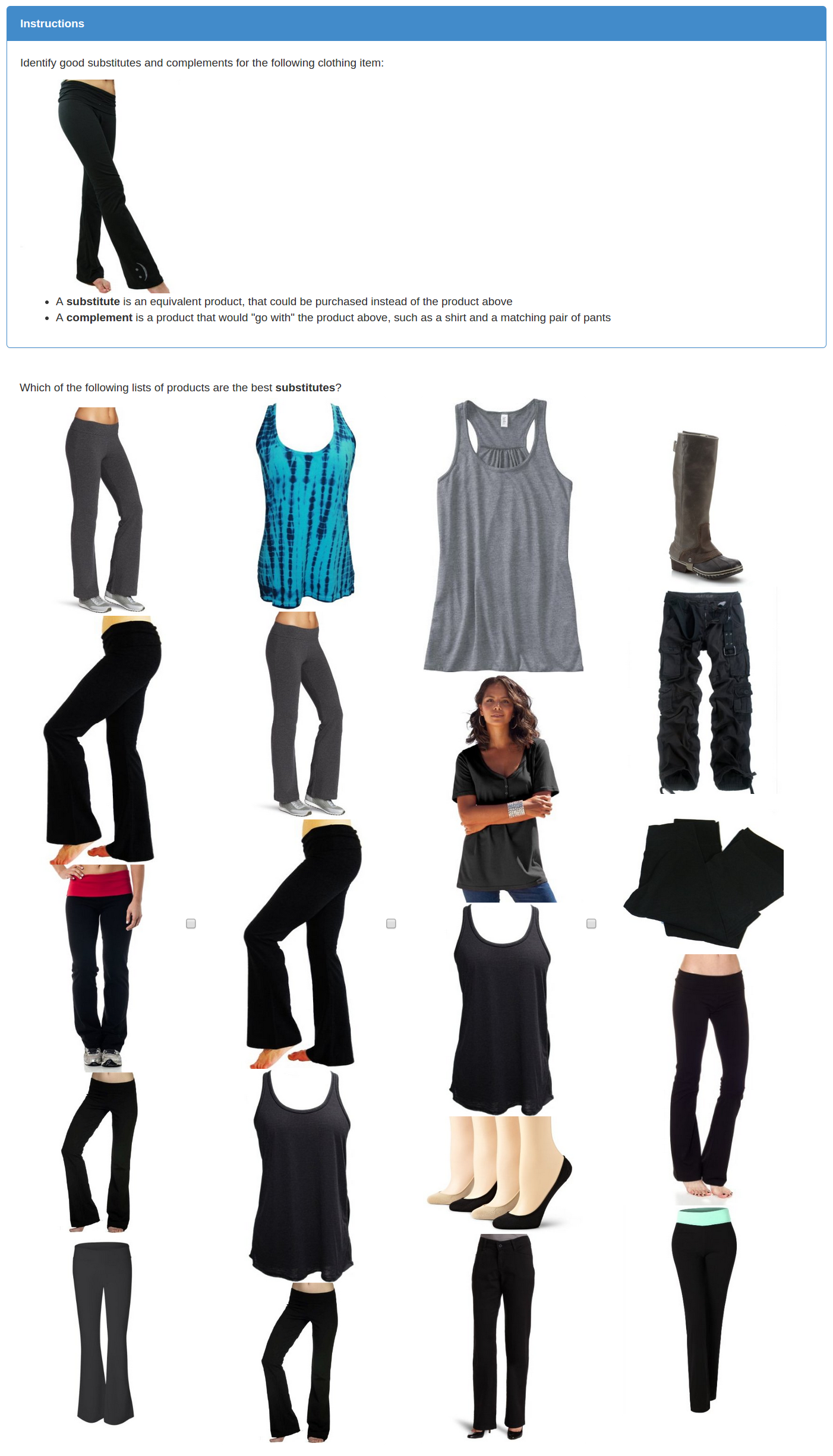}}}
}
\vspace{2mm}

\caption{(a,b,c) Examples of recommendations produced by \modelname{}; the top of each subfigure shows the query product, the left column shows substitutes recommended by \modelname{}, and the right column shows complements. (d) Interface used to evaluate \modelname{} on Mechanical Turk; Turkers are shown lists of items suggested by \emph{Amazon} (\ie, the ground-truth) and \modelname{} and must identify which lists they prefer.
}\label{fig:examples}
\end{figure*}

Having produced ranked lists of recommended relationships, our final task is to surface these recommendations to a potential user of our system in a meaningful way.

While conceptually simple, comparing all products against all others quickly becomes impractical in a corpus with millions of products. Our goal here is to rank all links, and surface those which have the highest likelihood under the model. That is, for each graph type $g$ we would like to recommend
\begin{equation}
 \mathit{rec}_g(i) = \argmax_{S \in (\mathcal T \setminus \lbrace i \rbrace)^R} \sum_{j \in S} F_{\beta_g}^\leftrightarrow(\psi_\theta(i,j)) F_{\eta_g}^\rightarrow(\varphi_\theta(i,j)),
\end{equation}
where $S \in (\mathcal T \setminus \lbrace i \rbrace)^R$ is a set of $R$ products other than $i$ itself.

While computing the score for a single candidate edge is very fast ($O(K)$ operations), on a dataset with millions of products this already results in an unacceptable delay when ranking all possible recommendations. 
Similar to \cite{Linden03} we implemented two modifications that make this enumeration procedure feasible (on the order of a few milliseconds). The first is to ignore obscure products by limiting the search space by some popularity threshold; we consider the hundred-thous\-and most popular products per-category when generating new recommendations. The second is to cull the search space using the category tree explicitly; \eg~when browsing for running shoes we can ignore, say, camera equipment and limit our search to clothing and shoes. Specifically, we only consider items belonging to the same category, its parent category, its child categories, and its sibling categories (in other words, its `immediate family'). It is very rare that the top-ranked recommendations belong to distant categories, so this has minimal impact on performance.

Another issue is that of adding new products to the system. Naturally, it is not feasible to re-train the system every time a new product is added. However, this is thankfully not necessary, as the introduction of a small number of products will not fundamentally change the word distribution $\phi$. Thus it is simply a matter of estimating the product's topic distribution under the existing model, as can be done using LDA \cite{blei_lda}.

When assembling our user interface (see Figs.~\ref{fig:1} and \ref{fig:examples}) we use the discovered topics from Section \ref{sec:microcategories} to `explain' recommendations to users, by selecting sentences whose language best explains why the recommended product was predicted. Specifically, we highlight sentences whose words yield the largest response to $F^{\rightarrow}_{\eta_g}$.

\xhdr{Reproducing Sceptre} All data and code used in this paper, as well as the interface from Figure \ref{fig:examples} is available on the first author's webpage: \url{http://cseweb.ucsd.edu/~jmcauley/}.

\section{Conclusion}

A useful recommender system must produce recommendations that not only match our preferences, but which are also \emph{relevant} to our current topic of interest. For a user browsing a particular product, two useful notions of relevant recommendations include \emph{substitutes} and \emph{complements}: products that can be purchased \emph{instead of} each other, and products that can be purchased \emph{in addition to} each other. In this paper, our goal has been to \emph{learn} these concepts from product features, especially from the text of their reviews.

We have presented \modelname{}, a model for predicting and understanding relationships between linked products. We have applied this to the problem of identifying substitutable and complementary products on a large collection of \emph{Amazon} data, including 144 million reviews and 237 million ground-truth relationships based on browsing and co-purchasing logs.

\small
\setlength{\bibsep}{2.5pt}

\end{document}